\documentclass[]{mn2e}
\usepackage{epsfig}
\newif\ifAMStwofonts

\title{The optical counterpart of the ultra-luminous x-ray source 
NGC~5204~X-1\thanks{\tiny{This work is based on observations made with the
NASA/ESA Hubble Space Telescope, obtained from the data archive at the
Space Telescope Science Institute. STScI is operated by the
Association of Universities for Research in Astronomy, Inc. under NASA
contract NAS 5-26555.}}}

\author[M.R. Goad et al.]
       {M.R. Goad$^{1}$, T.P. Roberts$^{2}$, C. Knigge$^{1}$, P. Lira$^{3}$ \\
        (1) University of Southampton, (2) University of Leicester (3) Universidad de Chile}
\date{Received 2002 July 2}

\pagerange{\pageref{firstpage}--\pageref{lastpage}}
\pubyear{2002}
\begin{document}

\maketitle

\label{firstpage}

\begin{abstract}

We use archival HST/WFPC2 V and I band images to show that the optical
counterpart to the ultra-luminous x-ray source NGC~5204 X-1, reported
by Roberts et~al., is composed of two sources separated by 0.5''. We
have also identified a third source as a possible counterpart, which
lies 0.8'' from the nominal X-ray position. PSF fitting photometry
yields V-band magnitudes of 20.3, 22.0 and 22.4 for the three sources.
The V$-$I band colours are $0.6$, $0.1$, and $-0.2$, respectively
(i.e. the fainter sources are bluer). We find that all V$-$I colours
and luminosities are consistent with those expected for young stellar
clusters (age $<10$~Myr).

\end{abstract}


\section{Introduction}

EINSTEIN and ROSAT observations show that at high energies most nearby
(spiral) galaxies are dominated by a relatively small number of
discrete, compact, but highly luminous x-ray sources.  These so-called
ultra-luminous x-ray sources (ULXs) have luminosities of $L_{\rm
x}\sim10^{39}-10^{41}$~erg~s$^{-1}$, far in excess of the Eddington
limit for spherical accretion onto a neutron star (Roberts and Warwick
2000).

Nearly 20\% of ULXs appear to be associated with supernova remnants
(SNR, Roberts et~al. 2002), while the rest are thought to be powered
by accretion onto a compact object. One possibility is that ULXs
represent the missing class of intermediate mass
($10^{2}-10^{5}$M$_{\odot}$) black holes (Colbert and Mushotzky
1999). However, recent Chandra observations reveal a strong
association between ULXs and active star-forming regions (Fabbiano
et~al. 2001, Lira et~al. 2002), that are far too young to contain
massive black holes (King et~al. 2001).  Subsequently, at least two
alternative scenarios have been proposed.  In the first, ULXs are
stellar mass black hole binaries undergoing a period of
super-Eddington accretion (e.g. Watarai et~al. 2001, Begelman
et~al. 2002).  In the second, the apparent super-Eddington
luminosities are due to the beamed x-ray emission from an otherwise
normal intermediate/high-mass x-ray binary (King et~al. 2001,
Georganopoulos et~al. 2002). While both scenarios tie in neatly with
the observed association between ULXs and active star forming regions
(Long et~al. 2002, Terashima and Wilson 2002, Zezas et~al. 2002), the
latter requires a significantly larger number of ULXs to be present.

Roberts et~al. (2001) identified the first optical counterpart to a
ULX, NGC~5204~X-1, based on its Chandra position and optical
multi-fibre spectroscopy.  The counterpart, located within
1'' of the nominal Chandra position, displays a blue
featureless continuum and is surrounded by an ionized bubble
of gas with a diameter of some 360~pc (Pakull and Mirioni 2002).
Here, we present archival HST/WFPC2 V and I band images of this source,
together with an improved analysis of the multi-fibre data. These data
allow us to place more robust constraints on the nature of the
counterpart, and thus obtain a deeper insight into the ULX phenomenon.

\begin{figure*}
\epsfig{width=6.8in,file=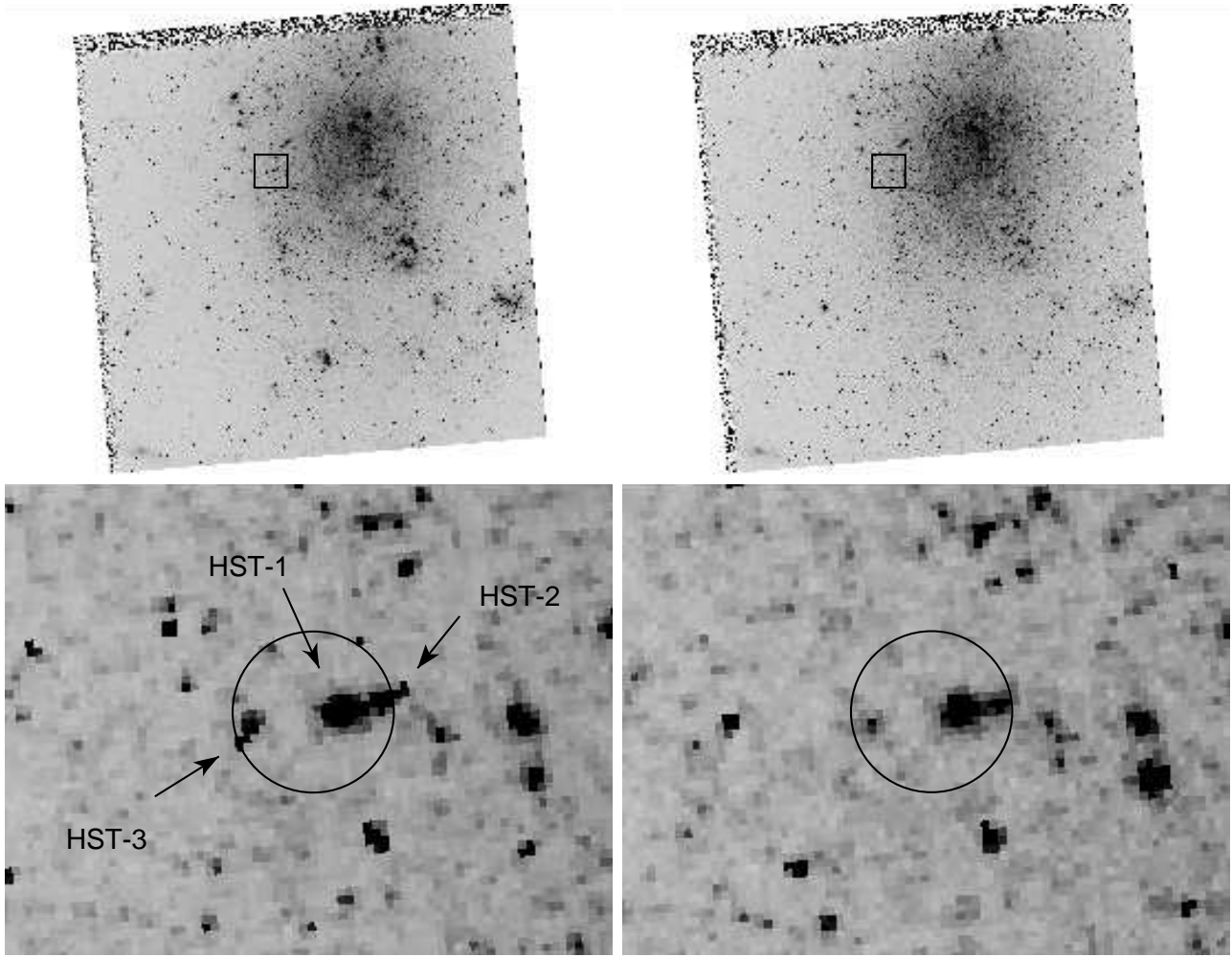}
\caption{Upper panel - HST WFPC2 V (left-hand side) and I 
(right-hand side) band images of NGC~5204 showing the location of the
optical counterpart. Lower panel - enlarged V and I band images reveal
that the optical counterpart is resolved into two components (HST-1
and HST-2) separated by $\sim$0.47''. A third source (HST-3) is
located 0.8'' east of the nominal Chandra position.  The 1.0'' radius
ring shows the formal error on the ULX position. North is up.}
\end{figure*}

\section[]{HST/WFPC2 images of NGC~5204~X-1}

We have obtained archival HST/WFPC2 V and I band (F606W and F814W)
images of the dwarf magellanic type galaxy NGC~5204 (Fig~1 - upper
panel). This data (now public) was obtained as part of an ongoing
HST/WFPC2 snapshot survey of nearby dwarf galaxies (Seitzer, PI).

The HST/WFPC2 V and I band images (Fig~1, lower panel) show that the
optical counterpart can be resolved into at least two sources, here
designated HST-1 and HST-2, separated by 0.47''.  We also identify a
third source designated HST-3, located 0.8'' to the east of the
nominal Chandra position. This source was not detected in the fibre
data of Roberts et~al. but due to its close proximity remains a
possible counterpart to the ULX.  The observed separations correspond
to distances of 10$-$16~pc at the distance of NGC~5204
(4.8~Mpc). 

We have performed PSF fitting photometry on the V and I band images
using HSTphot (Dolphin 2000).  Unfortunately, only one image was
available in each filter, hampering cosmic ray rejection.  Given the
difficulty of identifying real point sources, we decided against
iterative adjustment of HSTphot's standard PSFs on the basis of the
fit residuals.  The chip positions, RA, DEC, V and I band magnitudes
(referenced to a 0.5'' aperture), colours, and distances from the
nominal Chandra position ($\Delta$), are quoted in Table~1 for each
source.

\begin{table*}
 \centering
  \caption{HSTphot PSF fitting photometry}
  \begin{tabular}{@{}ccccccccrc@{}}
Source & Chip no. & \multicolumn{2}{c}{Chip position} & RA & DEC & V  &  I  & V$-$I & $\Delta$\\
     &   & X      & Y      &               &              &      &      &  & (arcsecs) \\
  HST-1  & 2 & 273.75 & 343.96 & 13:29:38.57 & 58:25:05.78 & 20.3 & 19.7 & 0.6 & 0.34\\
  HST-2  & 2 & 273.37 & 348.66 & 13:29:38.51 & 58:25:05.86 & 22.0 & 21.9 & 0.1 & 0.82\\
  HST-3  & 2 & 274.44 & 332.56 & 13:29:38.71 & 58:25:05.62 & 22.4 & 22.6 & $-$0.2 & 0.80 \\
\end{tabular}
\end{table*}

Our HSTphot photometry reveals that all sources are blue
(i.e. $F_{\lambda}$ decreases with increasing $\lambda$), consistent
with the finding of Roberts et.~al. for a single unresolved source.
However the two fainter sources are significantly bluer than HST-1.
The combined magnitudes of HST-1 and HST-2 are 20.1 in V and 19.6 in
I, with a V$-$I colour of 0.5.  These magnitudes are referenced to a
0.5'' aperture, becoming 20.0 and 19.5 when corrected to infinite
aperture. We do not include HST-3 in the combined magnitude
estimate, as it does not fall within the same fibre in the data of
Roberts et~al.

The best-fit Chandra position for the ULX (determined using WAVDETECT;
Freeman et~al. 2002) -- $\alpha=$13~29~38.61, $\delta=$+58~25~05.7 --
is located 0.34'' east of HST-1.  The 1$\sigma$ error associated with
the Chandra reference frame is $\pm 0.6$'' (Aldcroft
et~al. 2000). However, errors of up to 1.0'' are not uncommon when
linking the Chandra and HST reference frames (see e.g. Grindlay
et~al. 2001).  Thus while HST-1 is the preferred location for the
optical counterpart, neither HST-2 or HST-3 can be ruled out.

\begin{figure}
\epsfig{width=3.4in,height=3.5in,file=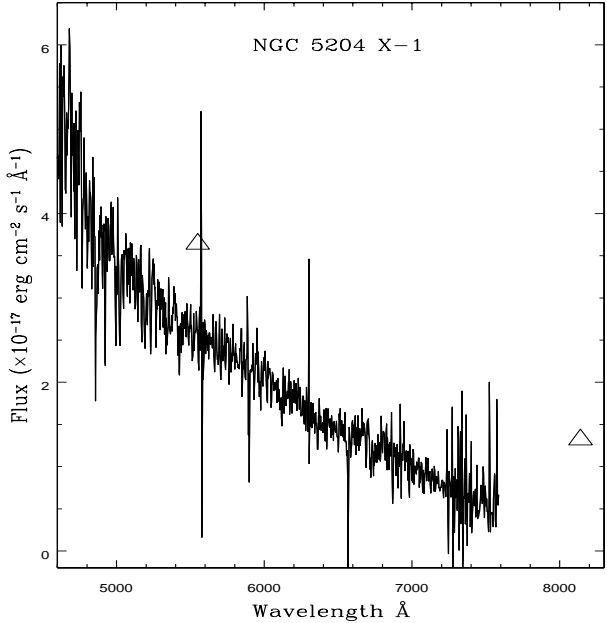} \caption{The optical
counterpart to NGC~5204 X-1 after removal of contaminating nightsky and host
galaxy emission. The open triangles represent the derived V and I
band fluxes for the combined sources as estimated from the HST/WFPC2
images.}
\end{figure}

\section[]{Comparison with WHT/INTEGRAL multi-fibre spectroscopy}

We have reanalysed the optical multi-fibre data on NGC~5204~X-1 from
Roberts et~al. (2001) with the aim of improving the photometric
accuracy of their spectrum. This allows a direct comparison of their
results with our new results from HSTphot. Our new analysis takes
better account of fibre throughput corrections (using an improved
throughput image, with repeatability errors of $<0.1$\%) and
contamination by the nightsky background. More importantly, we have
also tried to correct for diffuse emission from within the host
galaxy.

The reconstructed multi-fibre image of our source, shows that emission
from the optical counterpart lies within a single central fibre and
its surrounding nearest 6 neighbours. If we adopt the procedure
followed by Roberts et~al. and correct for nightsky background only,
we derive a V-band magnitude of 19.9. This is 0.2 magnitudes fainter
than found by Roberts et~al.; this difference is due partly to our
improved throughput correction and partly to the use of a median
(instead of a mean) nightsky background.

As noted above, we have also tried to account for diffuse emission
from the host galaxy, which shows a clear west-east gradient (see
Fig~1). We do this by taking a median of the local background
continuum over 18 fibres arranged in a hexagonal ring $\sim3$
arcseconds from the source.  This avoids including source flux in the
background fibres. As in Roberts et~al. the resultant spectrum (Fig~2)
is blue and featureless at the resolution of our instrument (6~\AA\
per 2 pixel resolution element).  However, removal of local galaxy
contributions results in a cleaner, fainter, and marginally bluer
spectrum. The measured flux near 5500~\AA\ translates to a V-band
magnitude of $20.4\pm0.1$. This is 0.7~mag fainter than the magnitude
quoted by Roberts et~al. and 0.4~mag fainter than that derived from
the HST/WFPC2 V-band image.  The discrepancy between the WHT V-band
flux estimate and that determined directly from the HST/WFPC2 V-band
image most likely results from the intermittent presence of high cloud
during some of the ground-based observations. This gives an indication
of the accuracy of the absolute spectral flux calibration of the fibre
data.




\section{The nature of the ULX and its counterpart.}

Roberts et~al. (2001) found that the optical counterpart to
NGC~5204~X-1 was unlikely to be a galactic foreground object, but
could not rule out the possibility that it was a background
BL~Lac. This was because the derived x-ray to optical slope
$\alpha_{\rm ox}\approx1.0$, lies well within the parameter range
occupied by BL~Lacs. However, a recent measurement of the radio-flux
at 8.3 GHz with the VLA (Wong et~al. 2002), shows the source to be
radio faint ($<~84\mu$J), yielding an upper limit to $\alpha_{\rm
ro}<$0.08. This places the source well outside the parameter space
occupied by BL~Lacs and confirms the status of NGC~5204~X-1 as a
bona-fide ULX.

\subsection{ULXs in young star clusters?}

The HST/WFPC2 colours show that our composite WHT spectrum, which is
dominated by the brighter and significantly redder source, must
flatten towards longer wavelengths. Based on the colours alone, HST-1
has an apparent spectral type F2-F5, HST-2 is type A2, and HST-3 type
B5-B6 (Zombeck 1990).  The absolute magnitudes of the three sources
are $M_{v}=-8.1$ (HST-1), $M_{v}=-6.4$ (HST-2), and $M_{v}=-6.0$
(HST-3), which at the bare minimum requires $\sim$1-2 F supergiants,
$\sim$2-3 A2 supergiants, and $\sim$2-3 B5 supergiants respectively.
However, a much more likely scenario is that the colours are
representative of young stellar clusters.
\begin{figure}
\epsfig{width=3.4in,height=4.0in,file=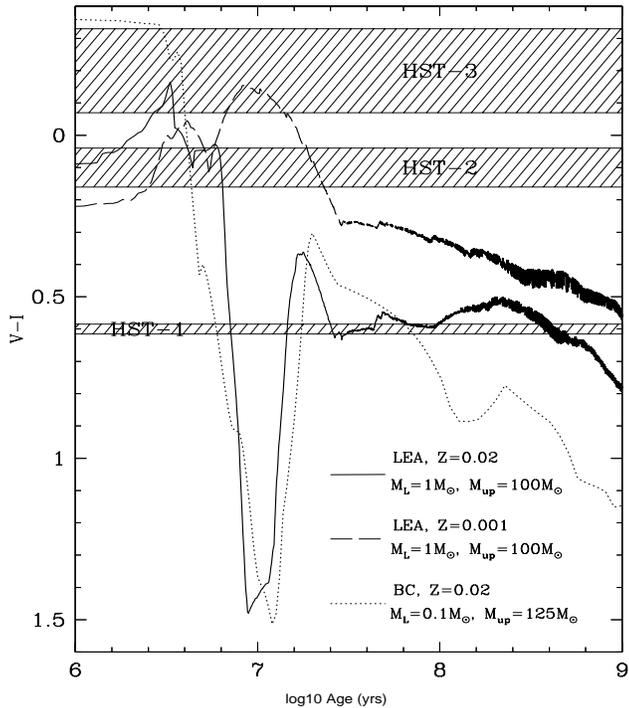}
\caption{V$-$I colours as a function of age for an instantaneous
burst of star formation (see text for details).  Individual models
represent : Starburst99 models of Leitherer et~al. (LEA) with solar
metallicity (solid line), 1/20th solar (dashed line), and the 1993
solar metallicity model of Bruzual and Charlot (BC, dotted line). The
shaded regions indicate the V$-$I colours of our three sources.}
\end{figure}
Indeed there is increasing evidence to support such an idea. Zezas
et~al.  (2002) find several instances of ULX/young stellar cluster
coincidences in the Antennae galaxies. Further, based on its location
and colour, one of the two optical counterparts to the known ULXs in
NGC~4565 has been associated with a globular cluster located in the
outer bulge of this galaxy (Wu et~al. 2002).

In Figure 4 we show V$-$I colours as a function of age for three stellar
cluster models. Each model assumes a single instantaneous burst of
star formation. The adopted stellar initial mass function (IMF) is
here represented by a powerlaw with slope 2.35 between the low- and
high-mass cut-offs and approximates the classical Salpeter IMF
(Salpter 1955), appropriate for most observations of star-forming
regions. Two of the models are from the Starburst99 models of
Leitherer et~al. (1999). For comparison, we also show the solar
metallicity model of Bruzual and Charlot (1993), which highlights the
effect of adopting a smaller lower mass cut-off.

If each of our sources represents a stellar cluster, and the clusters
are co-eval, then Fig~4 shows that the observed V$-$I band colours are
broadly consistent with young stellar clusters with ages between
$10^{6.6}-10^{7.3}$~yrs. Furthermore, asuming that the clusters have
the same metallicities, then the low-metallicity model is inconsistent
with the V$-$I colour of HST-1, and the age constraint is tightened to
between $10^{6.6}-10^{6.9}$~yrs.  This timescale is much shorter than
the formation timescale of a low mass x-ray binary,
$\sim 10^{8}-10^{9}$~yrs, but is consistent with the formation timescale
of an intermediate/high mass x-ray binary, $\sim10^{7}$~yrs (King
et~al. 2001).

By comparing the absolute V-band magnitudes, with the V-band
magnitudes predicted by the nuclear Starburst99 models of Leitherer et
al. (1999), we estimate the number of cluster members to be between
several hundred (HST-3) and a few thousand stars (HST-1). We note that
the upper limit to our source ``sizes'' (a few pcs at the distance of
NGC~5204) are consistent with cluster sizes determined for nearby
starburst galaxies (2-3 pcs, see Meurer et~al. 1995).

\section{Conclusions}

We have used archival HST/WFPC2 V and I band images to show that the
previously reported optical counterpart to the ultra-luminous x-ray
source NGC~5204 X-1, can be resolved into two sources separated by
0.47''. We have also identified a third possible counterpart 0.8''
east of the nominal Chandra position. All three sources appear blue
(i.e. $F_{\lambda}$ decreases with increasing $\lambda$), with the
fainter sources being significantly bluer. PSF fitting photometry with
HSTphot yields source magnitudes of 20.3, 22.0 and 22.4 in V.
Assuming that all three sources are co-eval, the V$-$I band colours
suggest that NGC~5204~X-1 lies within a young ($<$10~Myr) stellar
cluster.  This is consistent with the idea that ULXs are
intermediate/high-mass x-ray binaries.

\label{lastpage}

\end{document}